\documentclass[aaspp4]{aastex}
\usepackage{natbib,natbibspacing}
\usepackage{amsmath}
\usepackage{color}
\begin{document}
%

\title{Forecasting the Impact of Stellar Activity on Transiting Exoplanet Spectra}

\author{Robert T. Zellem$^{1}$, Mark R. Swain$^{1}$, Gael Roudier$^{1}$, Evgenya L. Shkolnik$^{2}$, Michelle J. Creech-Eakman$^{3}$, David R. Ciardi$^{4}$, Michael R. Line$^{2}$, Aishwarya R. Iyer$^{1}$, Geoffrey Bryden$^{1}$, Joe Llama$^{5}$, Kristen A. Fahy$^{1}$}

\affil{$^{1}$ Jet Propulsion Laboratory, California Institute of Technology, 4800 Oak Grove Drive, Pasadena, California 91109, USA}
\affil{$^{2}$ School of Earth \& Space Exploration, Arizona State University, Tempe, Arizona 85281, USA}
\affil{$^{3}$ Department of Physics, New Mexico Institute of Mining and Technology, 801 Leroy Place, Socorro, New Mexico  87801, USA}
\affil{$^{4}$ NASA Exoplanet Science Institute, California Institute of Technology, 1200 East California Blvd., Pasadena, CA 91125, USA}
\affil{$^{5}$ Lowell Observatory, 1400 West Mars Hill Road, Flagstaff, Arizona 86001, USA}

\email{rzellem@jpl.nasa.gov}

\vspace{12pt}
\begin{abstract}


Exoplanet host star activity, in the form of unocculted star spots or faculae, alters the observed transmission and emission spectra of the exoplanet. This effect can be exacerbated when combining data from different epochs if the stellar photosphere varies between observations due to activity. Here we present a method to characterize and correct for relative changes due to stellar activity by exploiting multi-epoch ($\ge2$~visits/transits) observations to place them in a consistent reference frame. Using measurements from portions of the planet's orbit where negligible planet transmission or emission can be assumed, we determine changes to the stellar spectral amplitude. With the analytical methods described here, we predict the impact of stellar variability on transit observations. Supplementing these forecasts with Kepler-measured stellar variabilities for F-, G-, K-, and M-dwarfs, and predicted transit precisions by JWST's NIRISS, NIRCam, and MIRI, we conclude that stellar activity does not impact infrared transiting exoplanet observations of most presently-known or predicted TESS targets by current or near-future platforms, such as JWST, as activity-induced spectral changes are below the measurement precision.

\end{abstract}

\vspace{12pt}
\section{Introduction}
Stellar variability, in the form of star spots and faculae, can affect the measured transit depth of an exoplanet and hence its spectrum and retrieved physical properties \citep{pont08, silvavalio08, czesla09, wolter09, agol10, berta11, carter11, desert11, sing11, fraine14, mccullough14, oshagh14, damasso15, barstow15, zellem15, rackham17}. As a worst case scenario for very active stars, unocculted spots can cause an underestimation of the planet-to-star radius ratio of up to 4\% at near-infrared wavelengths and  10\% at visible wavelengths while faculae can cause an overestimation of the planet-to-star radius ratio of up to $\sim$0.2\% at near-infrared wavelengths at $\sim$3\% at visible wavelengths \citep{oshagh14}. Unocculted spots can also mimic a Rayleigh scattering slope indicative of haze; for example, the visible and near-IR slope of the exoplanet HD~189733b's transit absorption spectrum, interpreted as Rayleigh scattering by haze particles \citep{pont08, pont13, sing11, sing16}, has also been interpreted as unocculted star spots on its active K0 host star \citep{mccullough14}. Unocculted spots can also introduce false molecular spectral modulation into an exoplanet's spectrum, such as H$_{2}$O if the spots are sufficiently cool \citep{fraine14, barstow15}. Stellar faculae, which are brighter than the stellar photosphere, decrease the transit depth at optical wavelengths \citep{rackham17}. Evolving unocculted spots on an active host star could also pose a problem when stitching data together from multiple epochs spanning multiple wavelengths, as will be required to completely sample an exoplanet from 0.6--28~$\mu$m with NASA's James Webb Space Telescope \citep[JWST; ][]{barstow15}. Therefore spectroscopic observations of an exoplanet orbiting an active star have the potential to result in an erroneous interpretation of its atmospheric properties, if the measurements have sufficient precision. As transit spectroscopic measurements become increasingly precise, especially in the JWST era, the possibility of contamination of the transit signal by star spots must be examined with care.

Fortunately current transit spectroscopy, for example with Hubble/WFC3, provides an opportunity to characterize exoplanet host stars as the out-of-transit and in-eclipse portions of the lightcurve probe emission from the host star alone, assuming negligible transmission and emission from the exoplanet. This method has the advantage that stellar activity can be quantified from the same dataset used to measure the planet's transit or eclipse, assuming sufficient phase coverage. Extensive amounts of high precision Hubble/WFC3 data are available for such an analysis \citep[e.g.,][]{sing16, iyer16, stevensonhst16, tsiaras17}, which is interesting for both directly assessing the potential unocculted stellar activity contamination of existing measurements and as an independent check on conclusions about stellar activity based on ground-based monitoring.

Previous studies have characterized the stellar activity of exoplanet host stars via long-term, ground-based, visible and near-infrared photometric \citep[e.g.,][]{berta11, desert11, knutson12, kreidberg15, narita13, nascimbeni15, pont08, pont13, sing11, zellem15} and spectroscopic monitoring \citep[e.g.,][]{baliunas95, lovis11, robertson13a, robertson13b}. However, ground-based observations are limited by the observability of the target during the year and weather, occasionally preventing them from being simultaneous with the space-based observations \citep[e.g.,][]{knutson12}. Non-contemporaneous observations make it difficult to determine the level of activity of the host star at the time of the space-based observations. Other studies have characterized changes in an exoplanet's transmission spectrum via detailed modeling of spectral modulation of the host star due to unocculted spots and plages \citep{oshagh14,rackham17}. Alternatively we demonstrate here a method that uses the out-of-transit data itself to characterize stellar activity and its impact on the observed transiting exoplanet spectrum between epochs of observations.

As an illustrative example, we study here in detail the sub-Neptune GJ~1214b. This transiting exoplanet is one of the best sources to apply our analysis techniques as it has a relatively bright host star (H-mag$=9.094$), providing high precision measurements (SNR~$\approx$~7000 per pixel-based spectral channel, per orbit),  and has been observed with 15~visits over $\sim$1~year (27 September 2012 -- 20 August 2013) with Hubble/WFC3 and the G141 grism (1.15--1.65~$\micron$), allowing us to search for stellar variability and assess its impact on the observed exoplanet spectrum. With multiple Hubble visits, we quantify relative changes in the host star's spectrum with time, sensitive to the presentation, formation, and evolution of star spots and faculae.

Using the analytical expressions described here, Kepler-measured F-, G-, K-, and M-dwarf variabilities \citep{ciardi17}, and the projected precisions of JWST's MIRI, NIRCam, and NIRISS \citep{greene16}, we find that stellar variability does not impact a majority of transit observations for both currently-known and projected Transiting Exoplanet Survey Satellite \citep[TESS;][]{sullivan15} targets. Thus, our study anticipates additional future, repeated high-precision spectroscopic transit measurements to observe stellar variations in exoplanet host stars such as with the New Mexico Exoplanet Spectroscopic Survey Instrument \citep[NESSI;][]{jurgenson10}, which will commence a survey of $\sim$30 transiting exoplanets on the 200~inch Hale Telescope at Palomar Observatory within the next year, ESA's Atmospheric Remote-sensing Exoplanet Large-survey (ARIEL), a proposed dedicated transiting exoplanet survey that will repeatedly measure the spectra ($\sim$2--8~$\micron$) of hundreds of exoplanets with multiple visits \citep{puig16, tinetti16}, and JWST, which will measure the infrared spectra of tens of transiting exoplanets \citep{cowan15, stevenson16}.

\section{The Impact of Unocculted Stellar Activity on Transit Spectroscopy}\label{sect:math}
Unocculted star spots and faculae affect the observed planetary transmission spectrum via the equation \citep[][]{mccullough14}:
\begin{align}
\delta_{active,\:transit} &= \underbrace{\bigg( \frac{R_{planet}}{R_{star}} \bigg)^{2}}_{\delta_{quiescent,\:transit}} \cdot \bigg[ 1 + \bigg( \frac{R_{feature}}{R_{star}} \bigg)^{2} \bigg( \frac{B_{feature}}{B_{star}} - 1 \bigg) \bigg]^{-1}
\label{eqn:unoccultedspots}
\end{align}
where $\delta_{active,\:transit}$ is the observed transmission spectrum modulated by unocculted stellar activity (spots or faculae), $\delta_{quiescent,\:transit}$ is the observed transmission spectrum when the star is in a quiescent state, $R_{planet}$ and $R_{star}$ are the radii of the exoplanet and host star, respectively, and $B_{star}$ is the spectrum of the star. $R_{feature}$ is the effective radius of the features (spots or faculae), if all features visible on the surface of the star were to be combined into one, large complex. $B_{feature}$ is the spectrum of this complex. Please note that all of these variables are functions of wavelength; we have omitted the wavelength symbol $\lambda$ for clarity.

The impact of stellar activity on eclipse measurements is derived as follows. For an inactive star, the measured ``quiescent'' eclipse depth is:
\begin{align}
\delta_{quiescent,\:eclipse} &= \frac{F_{out\text{-}of\text{-}eclipse} - F_{in\text{-}eclipse}}{F_{in\text{-}eclipse}} \nonumber \\
&= \frac{F_{in\text{-}eclipse} + B_{planet}R_{planet}^{2} - F_{in\text{-}eclipse}}{F_{in\text{-}eclipse}} \nonumber \\
&= \frac{B_{planet}R_{planet}^{2}}{B_{star}R_{star}^{2}}.
\end{align}
Whereas for an active star, the observed eclipse depth is:
\begin{align}
\delta_{active,\:eclipse} &= \frac{F_{out\text{-}of\text{-}eclipse} - F_{in\text{-}eclipse}}{F_{in\text{-}eclipse}} \nonumber \\
&= \frac{F_{in\text{-}eclipse} + B_{planet}R_{planet}^{2} - F_{in\text{-}eclipse}}{F_{in\text{-}eclipse}} \nonumber \\
&= \frac{B_{planet}R_{planet}^{2}}{B_{star}R_{star}^{2} + B_{feature}R_{feature}^{2} - B_{star}R_{feature}^{2}} \nonumber \\
&= \frac{B_{planet}R_{planet}^{2}}{B_{star}R_{star}^{2} \cdot \bigg(1 + \frac{B_{feature}R_{feature}^{2}}{B_{star}R_{star}^{2}} - \frac{B_{star}R_{feature}^{2}}{B_{star}R_{star}^{2}}\bigg)} \nonumber \\
&= \underbrace{\frac{B_{planet}R_{planet}^{2}}{B_{star}R_{star}^{2}}}_{\delta_{quiescent,\:eclipse}} \cdot \bigg[ 1 + \bigg( \frac{R_{feature}}{R_{star}} \bigg)^{2} \bigg( \frac{B_{feature}}{B_{star}} - 1 \bigg) \bigg]^{-1}.
\label{eqn:unoccultedspots_eclipse}
\end{align}
Note that the equation for the active star eclipse depth $\delta_{active,\:eclipse}$ (Eqn.~\ref{eqn:unoccultedspots_eclipse}) is the same as for transit $\delta_{active,\:transit}$ (Eqn.~\ref{eqn:unoccultedspots}).

We then isolate and define the ``stellar activity correction factor'' $\zeta$, a wavelength-dependent function that describes how much unocculted spots and faculae alter the planetary transit depths from their true, inactive values:
\begin{align}
\zeta = 1 + \bigg( \frac{R_{feature}}{R_{star}} \bigg)^{2} \bigg( \frac{B_{feature}}{B_{star}} - 1 \bigg).
\label{eqn:ACF}
\end{align}
We can then rearrange Equation~\ref{eqn:ACF}:
\begin{align}
\zeta &= 1 + \bigg( \frac{R_{feature}}{R_{star}} \bigg)^{2} \bigg( \frac{B_{feature}}{B_{star}} - 1 \bigg) \nonumber \\
&= \frac{1}{B_{star}R_{star}^{2}} \bigg[ B_{star}R_{star}^{2} + B_{star}R_{feature}^{2} \bigg(\frac{B_{feature}}{B_{star}} - 1 \bigg) \bigg] \nonumber \\
&= \frac{1}{B_{star}R_{star}^{2}} [ B_{star}R_{star}^{2} + B_{feature}R_{feature}^{2} - B_{star}R_{feature}^{2} ] \nonumber \\
&= \frac{F_{active}}{F_{quiescent}}
\label{eqn:ACF_redux}
\end{align}
Where $F_{active}$ is the measured spectrum of the active star and $F_{quiescent}$ is the spectrum of the quiescent star.

By combining Equations~\ref{eqn:unoccultedspots} and \ref{eqn:unoccultedspots_eclipse} with \ref{eqn:ACF} and \ref{eqn:ACF_redux}, we find:
\begin{eqnarray}
\delta_{active} = \delta_{quiescent} \cdot \zeta^{-1} \nonumber \\
\delta_{active} = \delta_{quiescent} \frac{F_{quiescent}}{F_{active}} \nonumber \\
\frac{\delta_{active}}{\delta_{quiescent}} = \frac{F_{quiescent}}{F_{active}}.
\label{eqn:spot_eqn_simplified}
\end{eqnarray}
Thus the measured transit/eclipse depth is inversely proportional to the brightness of the star. For example, if the star decreases in brightness by 1\% due to an unocculted spot, then the transit/eclipse depth will deepen by 1\%, and vice versa. Thus by observing the spectrum of the star when it is quiescent $F_{quiescent}$, one can use Equation~\ref{eqn:spot_eqn_simplified} to correct the observed exoplanet's transmission or emission spectrum for unocculted stellar activity.

This equation is powerful as it allows one to observe exoplanets of interest that have the misfortune of having a variable host star. By observing the system at least twice, one can measure the spectra of the host star $F$ with the out-of-transit and in-eclipse portions of the lightcurve to quantify $\zeta$ (Section~\ref{sect:1214_analysis}) and then correct for the relative changes in the observed planetary spectra due to unocculted stellar activity with Equation~\ref{eqn:spot_eqn_simplified} to place the observations in a consistent reference frame.

However, with a limited number of repeat visits, determining which epoch(s) are active or quiescent is not necessarily straightforward. One could implement stellar models \citep[e.g., PHOENIX;][]{husser13} to determine which epoch(s) agree with the stellar model, indicating quiescence $F_{quiescent}$. Alternatively, one could use ground-based monitoring or multiple visits (e.g., GJ~1214 features 15~visits with Hubble/WFC3; see Section~\ref{sect:1214}) to monitor stellar changes. At the very least, one can use the measured stellar variability either with the data itself (Section~\ref{sect:1214_analysis}), with previous observations of typical stellar activities \citep[Section~\ref{sect:forecast}; e.g.,][]{ciardi17}, or with ground-based monitoring to estimate the change in the exoplanet's observed transit depth $\Delta\delta$:
\begin{align}
\Delta\delta &= \delta_{active} - \delta_{quiescent} \nonumber \\
&= \delta_{quiescent} \frac{F_{quiescent}}{F_{active}} - \delta_{quiescent} \nonumber \\
&= \delta_{quiescent} \bigg(\frac{F_{quiescent}}{F_{active}} - 1\bigg) \nonumber \\
&= \delta_{quiescent} \bigg(\frac{F_{quiescent} - F_{active}}{F_{active}}\bigg).
\label{eqn:hotjupiter_example}
\end{align}
Thus, for example, if a planet with a 1\% transit depth orbits a star that is active at the 0.1\% level, then the planet's observed transit depth will change by 10~ppm. Therefore, by knowing the typical activity of a host star, one can predict the relative change in the observed transit depth $\Delta\delta$ to determine if it is larger than the precision of their data and adversely impact their observations (Section~\ref{sect:forecast}). However, if these additional sources of information are unavailable, then one can at the very least place all of the transit observations into a consistent reference frame.

\section{GJ 1214: An Example Variable System in the Hubble/WFC3 Bandpass}\label{sect:1214}

We demonstrate this stellar characterization method with 14~visits of Hubble/WFC3 spectroscopy ($1.15-1.65$~$\micron$) of GJ~1214 spanning 27 September 2012 -- 20 August 2013 (PID: 13021, PI: Jacob Bean). These data form the basis for the published spectrum of the sub-Neptune GJ~1214b, which was found to be flat, suggesting high-altitude clouds \citep{kreidberg14}. The exoplanet host star GJ~1214 is a nearby \citep[14.6~pc = 47.5~ly;][]{angladaescude13} M4.5V dwarf star \citep{lurie14}. It is metal rich, with a spectroscopically-derived [Fe/H] ranging from 0.26 to 0.39 \citep{angladaescude13, newton15, rojasayala10}. GJ~1214 is 3--10~Gyr old \citep{reidgizis05,charbonneau09}; at these older ages, stellar activity is expected to be fairly low, which is generally consistent with Hubble/STIS observations of the system that show little to no Lyman~$\alpha$ emission \citep{france13}.

Despite this finding, multiple studies have determined that GJ~1214 is active due to both star spots and faculae \citep{berta11, carter11, charbonneau09, fraine13, kreidberg14, kundurthy11, rackham17}. GJ~1214's variability is not unexpected: for those M-dwarfs that are located $\le$25~pc away, 50\%$\pm$25\% of M4 stars are active for 4.5$^{+0.5}_{-1.0}$~Gyr and 90\%$^{+2.5\%}_{-17.5\%}$ of M5 stars are active for 7.0$\pm$0.5~Gyr \citep{west08}. As such, multiple groups have conducted large ground-based photometric campaigns to characterize GJ~1214's long-term stellar variability. For example, the MEarth survey (40~cm telescope; Sloan $i+z$ photometry) found that GJ~1214 was active during the spring observing seasons in 2008, 2009, and 2010 with a semi-amplitude $A = 3.5 \pm 0.7$~mmag and a period of 53~days and V-band photometry with the 1.2~m FLWO/KeplerCam at Mt. Hopkins, AZ, from 2010 March 26 to 2010 June 17 measured $A = 7 \pm 3$~mmag and a period of 41~days \citep{berta11}. A large 32~night photometric campaign spanning 78~nights in 2012 with the 50~cm MISuME telescope determined that GJ~1214 was active with a period of $44.3 \pm 1.2$~days with $A = 3.47 \pm 0.41$~mmag in the $I_{c}$ band, $A = 0.61 \pm 0.04$~mmag in the $R_{c}$ band, and $A = 2.34 \pm 0.46$~mmag in the $g'$ band \citep{narita13}. A 191~night multi-band photometric campaign (110~nights from 21 March to 10 October 2012 and 81~nights from 24 March to 4 October 2013) with the 1.2~m STELLA/WiFSIP found that GJ~1214 was active in the $V_{J}$ band from 2012--2013 with $A = 11.4 \pm 1.1$~mmag  and a period of $83.0 \pm 0.8$~days, in the $V_{J}$ band in 2012 with $A = 15.1 \pm 1.3$~mmag and a period of $69.0 \pm 2.0$~days, in the $I_{c}$ band in 2012 with $A = 12.0 \pm 1.2$~mmag and a period of $79.6 \pm 2.5$~days, and in the $B_{J}$ band in 2012 with $A \ge 15$~mmag and a period of $\ge70$~days \citep{nascimbeni15}. For comparison, based on a study of Kepler observations (0.42--0.9~$\micron$), M-dwarfs have a median activity level of 6.83~mmag over a baseline of 100~days \citep{ciardi17}. Thus compared to other M-dwarfs, GJ~1214's variability is typical.






To further investigate stellar variability with a heretofore unused resource, we characterize GJ~1214's stellar activity with the out-of-transit portions of the Hubble/WFC3 data itself. Then, using the formulae previously defined in Section~\ref{sect:math}, we quantify the effects of GJ~1214's stellar variability on the exoplanet's measured transmission spectrum.

\subsection{Data Extraction and Analysis}\label{sect:1214_analysis}

\begin{figure}[!htb]
\centering
\includegraphics[width=1\columnwidth]{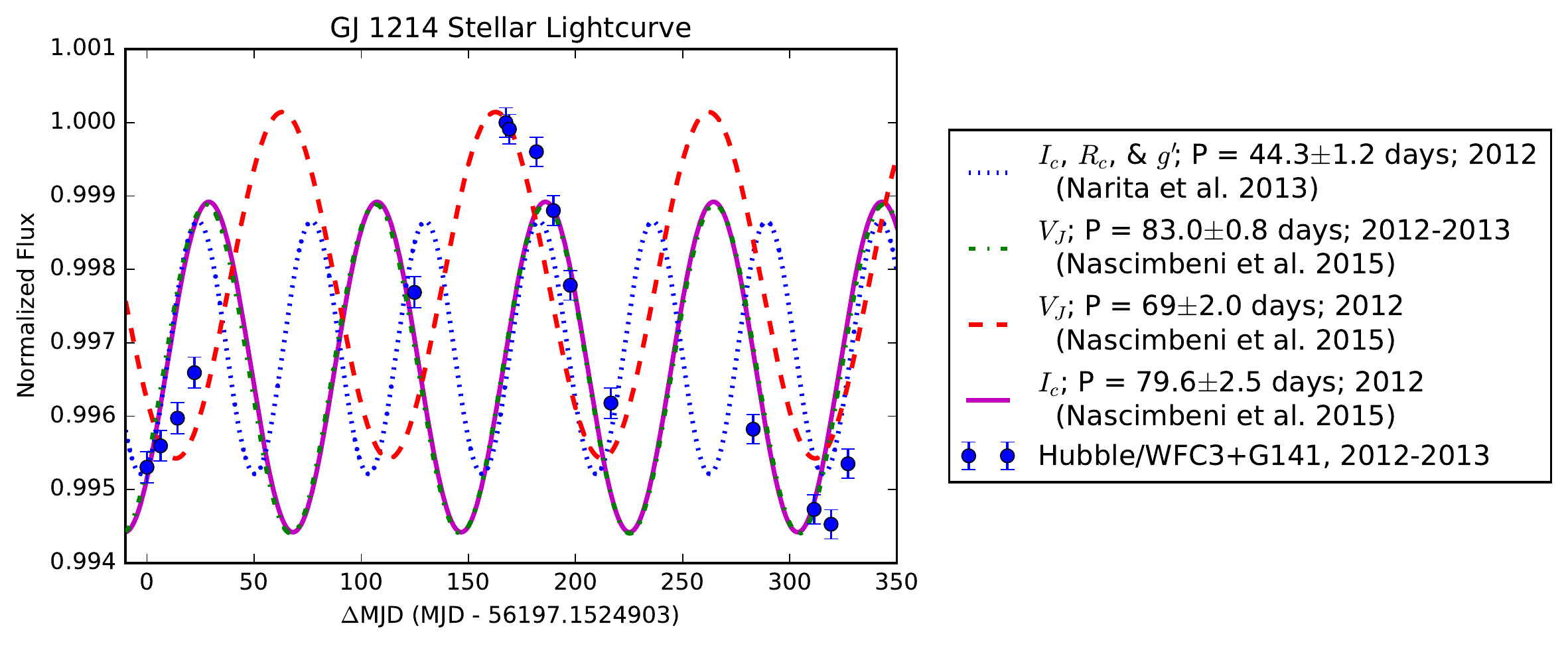}
\caption{\textbf{GJ 1214: An Exoplanet Host Star with Amplitude Variability.} GJ~1214's stellar lightcurve (blue circles), integrated over the Hubble/WFC3+G141 passband, indicates a semi-amplitude variability of $A = 0.279\% \pm 0.012\% = 2.34 \pm 0.46$~mmag, which is 3.77$\times$ smaller than semi-amplitudes measured in the optical over a similar time-frame \citep{nascimbeni15}. In addition, the near-infrared period $P$ of GJ~1214's variability, when fitting for the amplitude and phase offset, is incompatabile with previously-reported values measured at shorter wavelengths (dotted, dash-dot, dashed, and solid lines), potentially due to evolution of the activity.
}
\label{fig:GJ1214_lightcurve}
\end{figure}

We extract GJ~1214's time-varying Hubble/WFC3 spectra from the STScI-calibrated \texttt{ima} files \citep{dressel17} by taking the difference between non-destructive reads to provide a background subtraction, wavelength-calibrating the spectra with a G141 filter transmission curve, flagging bad pixels, and taking the mean along the spatial scan axis. We then remove the first orbit of each visit to mitigate Hubble/WFC3's ``ramp effect'' \citep{berta12, swain13, kreidberg14, kreidberg14b, stevenson14}, stack and order the spectra in time, isolate the out-of-transit portions, and sum over all wavelengths to construct GJ~1214's broadband stellar lightcurve (Fig.~\ref{fig:GJ1214_lightcurve}).


We find that GJ~1214 has a near-infrared semi-amplitude $A = 0.279\% \pm 0.012\% = 2.34 \pm 0.46$~mmag, which is 3.77$\times$ smaller than its visible wavelength ($V_{J}$ photometric band) semi-amplitude of $11.4 \pm 1.1$~mmag measured over a similar time-frame \citep{nascimbeni15}. We then fit GJ~1214's time-varying flux with multiple sine waves using a Markov Chain Monte Carlo \citep[MCMC; e.g.,][]{ford05} and adopting as priors previously-reported periods ($44.3 \pm 1.2$~days, \citet{narita13}; $83.0 \pm 0.8$, $69.0 \pm 2.0$, and $79.6 \pm 2.5$~days, \citet{nascimbeni15}). However none of these periods, measured at shorter wavelengths, fit all of our near-infrared data (Fig.~\ref{fig:GJ1214_lightcurve}), including the period of $83.0 \pm 0.8$~days, measured in the $V_{J}$ band over a similar time-frame as the Hubble/WFC3 observations \citep{nascimbeni15}. This discrepancy is potentially due to the fact that these visible-band data (21 March -- 10 October 2012; 24 March -- 4 October 2013) were not fully contemporaneous with the Hubble/WFC3 observations (27 September -- 19 October 2012; 30 January -- 1 May 2013; 7 July -- 20 August 2013). Thus GJ~1214's spots or faculae could have evolved on its surface, changing its apparent period, thereby rendering our near-infrared result incompatible with its visible-band period. However, despite this apparent spot evolution, our data agree and complement to first order GJ~1214's 2013 $V_{J}$ and $B_{J}$ photometry \citep{nascimbeni15}.

\subsection{Assessing the Impact of Stellar Activity on Planetary Spectra}
We next quantify the effect of this stellar activity on the observed per-visit GJ~1214b spectra. Because only a final, visit-averaged spectrum was published for GJ~1214b \citep{kreidberg14}, we first estimate the planet's per-visit spectra by increasing the uncertainties on the published spectra by the square root of the number of visits (14). Since we explore how much the planetary spectrum could change due to unocculted activity, we invert Equation~\ref{eqn:spot_eqn_simplified} to solve for the active-star planetary spectrum $\delta_{active}$ for each visit and use the mean stellar spectrum as a reference in place of GJ~1214's quiescent spectrum $\delta_{quiescent}$.

\begin{figure}[!htb] 
\centering
\includegraphics[width=0.6\columnwidth]{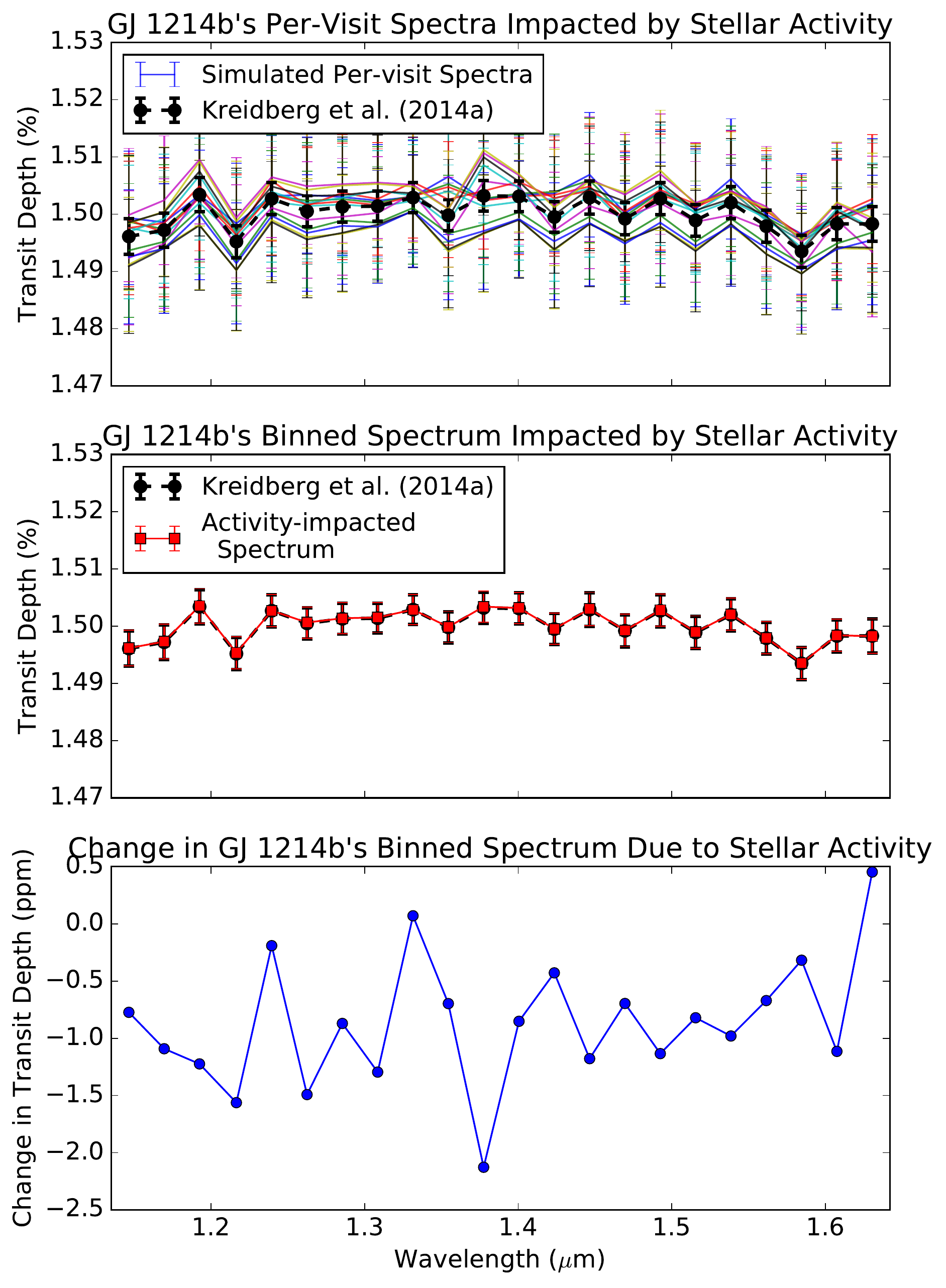}    
\caption{\textbf{Stellar Activity Does Not Affect GJ~1214b's Spectrum.} \textit{Top Panel:} The visit-to-visit changes in the exoplanet's measured transmission spectrum due to stellar activity (thin, multi-colored lines) compared to GJ~1214b's visit-binned published spectra \citep[black circles;][]{kreidberg14}. \textit{Middle Panel:} Transmission spectrum constructed by averaging the stellar activity modulated per-visit spectra (red squares) compared to the published transmission spectrum (black circles). \textit{Bottom Panel:} Differential spectrum of the published transmission spectrum and the stellar activity modulated spectrum. Stellar activity does not statistically alter the exoplanet's measured spectrum.}
\label{fig:4panel}
\end{figure}

We find that for most spectral channels, stellar activity does not measurably modulate the planet's per-visit spectrum (Fig.~\ref{fig:4panel}, Top). This outcome is even more clear when we bin the per-visit spectra in time (Fig.~\ref{fig:4panel}, Middle and Bottom). Therefore, merely using only the Hubble/WFC3 data themselves, we confirm the assessment of \citet{kreidberg14}, which was based on simulations of stellar activity, that GJ~1214b's spectrum is not statistically impacted by stellar variability at this level of measurement precision ($\sim$30~ppm).

\section{Forecasting the Effect of Stellar Activity on Other Instruments} \label{sect:forecast}

\begin{figure}[!htb]
\centering
\includegraphics[width=1\columnwidth]{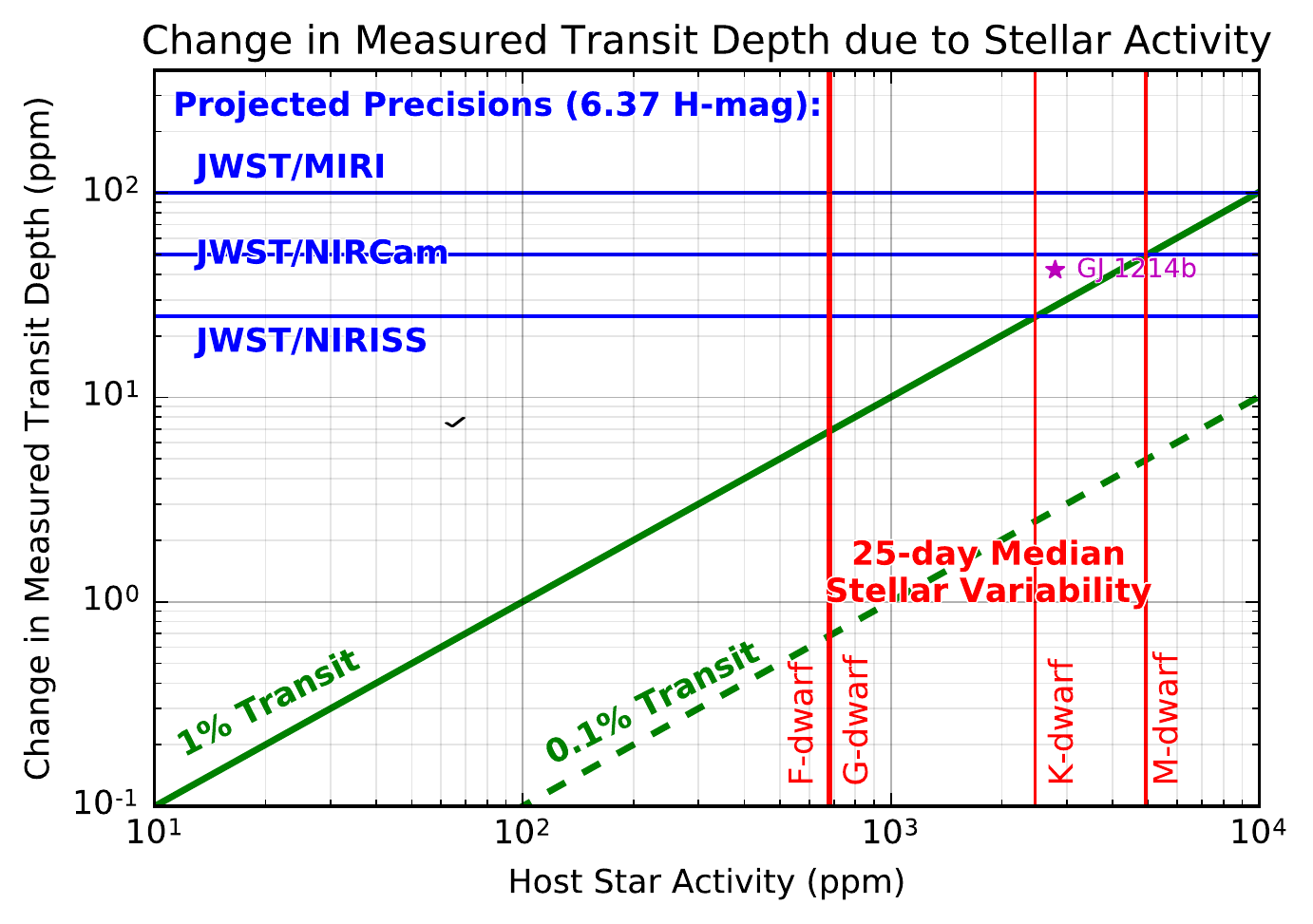}
\caption{\textbf{Host Star Activity Minimally Impacts Infrared Transit Observations.} Forecasted change in the exoplanet's measured transit depth as a function of its transit depth (green diagonal lines; Eqn.~\ref{eqn:hotjupiter_example}) and host star's activity. Also shown are the 25-day median stellar variabilities of F-, G-, K-, and M-dwarf stars as measured by Kepler \citep[red vertical lines;][]{ciardi17}; these variabilities provide a conservatively-high estimate as the effects of stellar activity on transit observations are more pronounced in the visible than in the infrared \citep{oshagh14,rackham17}. Also included are the projected transit precisions of JWST's NIRISS, NIRCam, and MIRI observations of an exoplanet with a 6.37~H-mag host star \citep[blue horizontal lines;][]{greene16} and GJ~1214's stellar variability measured in this study (magenta star; Fig.~\ref{fig:GJ1214_lightcurve}). We predict here that infrared observations by JWST, and therefore most present and near-future platforms, will largely be unaffected by stellar variability.}
\label{fig:activity_vs_depth}
\end{figure}

Using Equation~\ref{eqn:hotjupiter_example}, we forecast the change in an exoplanet's observed transit depth as a function of its transit depth and host star activity (Fig.~\ref{fig:activity_vs_depth}). Our measurement of GJ~1214's variability (magenta star) is included for reference. We also plot median stellar variabilities measured by Kepler over a 25-day period for F-, G-, K-, and M-dwarfs \citep{ciardi17}. We chose a 25-day timescale as it is similar to the amount of time TESS will observe each of its fields \citep[27.4~days; ][]{ricker14}. Thus one can use the stellar variability measured by TESS and Figure~\ref{fig:activity_vs_depth} to forecast if their transit measurements will be impacted by stellar activity. However since mid to late M-dwarf stars typically have longer rotation periods \citep[$\sim$100~days;][]{newton16}, this 25-day timescale may underestimate their variability. When using a 100-day timescale, we find that M-dwarfs have a median variability that is 1.3$\times$ larger than the 25-day timescale \citep{ciardi17}. We have omitted plotting the M-dwarf 100-day median variability on Figure~\ref{fig:activity_vs_depth} for clarity.

Stellar activity at visible wavelengths measured by Kepler are important because they provide a ``worst-case'' scenario for infrared transit measurements as stellar activity has a larger impact on the observed transit depth in the visible than the infrared \citep{oshagh14,rackham17}. Therefore, using Figure~\ref{fig:activity_vs_depth}, one can forecast the impact of stellar variability of their infrared observations on any platform given their target's transit depth and host star spectral type.

\begin{figure}[!htb]
\centering
\includegraphics[width=1\columnwidth]{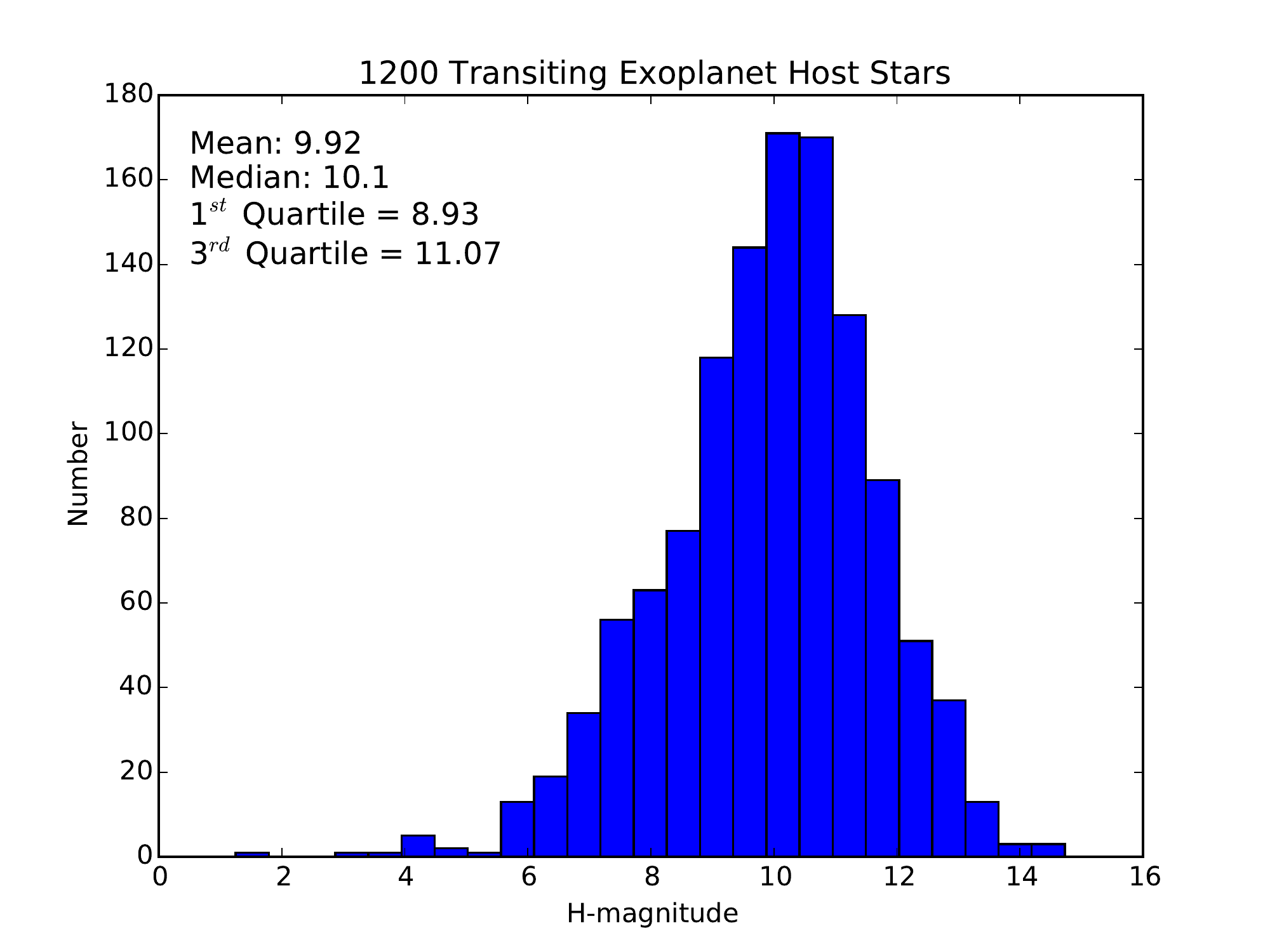}
\caption{\textbf{The 1200 Best Targets for Transiting Exoplanet Absorption Spectroscopy.} These targets, identified with a transit figure of merit (Eqn.~\ref{eqn:transit_FOM}), contain both presently-known targets and forecasted targets discovered by TESS \citep{sullivan15} and therefore represent the most likely targets of, for example, JWST.}
\label{fig:1200targets}
\end{figure}

We also include the projected measurement precisions of a 6.37~H-mag transiting exoplanet system with JWST's NIRISS, NIRCam, and MIRI \citep{greene16}. This magnitude limit is appropriate for guiding the next few years of transiting exoplanet observations as nearly all of the top 1200 targets are dimmer than 6~H-mag (Fig.~\ref{fig:1200targets}), when including both currently-known and predicted TESS yields \citep{sullivan15} ranked by a platform-independent figure of merit (FOM):
\begin{align}
\mathrm{FOM} &= \frac{\mathrm{signal}}{\mathrm{noise}} \\
&= \frac{\mathrm{1 \ H_{s} \ of \ spectral \ modulation}}{\mathrm{photon \ noise}} \\
&= \frac{2H_{s}R_{p}R_{s}^{-2}}{10^{0.2 H\text{-}mag}}
\label{eqn:transit_FOM}
\end{align}
where $H$-$mag$ is the host star's apparent magnitude in the H-band and $H_{s}$ is an exoplanet's scale height:
\begin{align}
H_{s} = \frac{k_{B}T_{eq}}{\mu g}
\end{align}
where $k_{B}$ is the Boltzmann constant, $T_{eq}$ is the planet's equilibrium temperature, $\mu$ is the planet's mean molecular weight, and $g$ is the planet's acceleration due to gravity. The values for $T_{eq}$ and $g$ are derived from parameters published in NASA's Exoplanet Archive \citep{akeson13}. However, since most of the targets in the 1200~planet sample lack detailed spectroscopic measurements, their atmospheric compositions  are unknown and therefore must be selected randomly from some prescribed distribution to estimate their mean molecular weight $\mu$.  Therefore we assign a metallicity to each planet based on an assumed relationship where metallicity increases as planet mass decreases. This relationship
is motivated both by simulations of planet formation \citep{fortney13} and by observations of methane in the Solar System's giant planets \citep{wong04, fletcher09, karkoschka11, sromovsky11}. The planet and star in each
system are assigned the same C/O ratio, drawn from a Gaussian distribution with a mean log[C/O] of -0.2 and a dispersion of 0.12 (cf. -0.26 for the Sun). Our choice for the C/O
distribution results in 5\% of systems having C/O~$>1$, which is within the range of
estimates from stellar observations \citep{fortney12, hinkel14}. Given the metallicity and C/O ratio assigned to each planet, we calculate the full range of molecular abundances within its
atmosphere assuming equilibrium chemistry with
the CEA code \citep{mcbride96}.
We then translate these abundances into an overall mean molecular
weight $\mu$ for each atmosphere.


We find that most infrared transit observations by JWST \citep{cowan15, stevenson16} and other current and near-future platforms, such as NESSI \citep{jurgenson10} and ARIEL \citep{puig16}, are unaffected by stellar variability, except for rare cases where the target is bright, relatively active, and has a large transit depth (e.g., HD~189733b or, alternatively, targets with a $\sim$1\% transit depth orbiting bright M-dwarfs). This conclusion is qualitatively in agreement with a study of the effects of stellar variability on the atmospheric retrievals of JWST transit spectroscopy observed at different epochs and observing modes \citep{barstow15}.

\section{Conclusions and Future Work}
Here we analytically derive the change in an exoplanet's transit depth as a function of host star activity and provide a method for mitigating stellar activity on multi-epoch ($\ge2$~visits) transit observations by characterizing the activity of the host star from the out-of-transit and in-eclipse portions of the lightcurve. An advantage of this technique is that it does not require modeling of the stellar activity itself to place them in a consistent reference frame, however incorporating unocculted spot and faculae modeling \citep{oshagh14,rackham17} in conjunction with the techniques presented here would be beneficial, as it has the potential to identify the star's quiescent state, and is subject to future work. Using Kepler observations of stellar variability for a variety of stellar spectral types and predicted precisions for JWST instruments, we conclude that infrared observations of currently-known and TESS-discovered transiting exoplanets by present and near-future platforms, such as JWST, are largely unaffected by stellar variability. However, certain highly-variable stars (e.g., HD~189733 or bright M-dwarfs with large ($\gtrsim1\%$) transit depths) may still need special consideration and a first order spectral correction can be implemented with the formalisms presented here (Section~\ref{sect:math}).

Using the methods provided here, we analyze publicly-available Hubble/WFC3 transit spectroscopy of GJ~1214 and determine that it is active with a semi-amplitude of $0.279\% \pm 0.012\% = 3.02 \pm 1.30$~mmag, 3.77$\times$ smaller than its measured activity at visible wavelengths over a similar time \citep[$11.4 \pm 1.1$~mmag;][]{nascimbeni15}. We also find that GJ~1214's visible-band periods are incompatible with its near-infrared data, potentially due to stellar activity evolution occuring between the observations. Regardless, we confirm that GJ~1214's stellar activity does not statistically impact GJ~1214b's Hubble/WFC3+G141 near-infrared spectrum.

\section*{Acknowledgments}
Part of the research was carried out at the Jet Propulsion Laboratory, California Institute of Technology, under contract with the National Aeronautics and Space Administration. Copyright 2017. All rights reserved.

R.T.Z. would like to thank JPL's ExoSpec team and Jacob Bean for their insightful comments.

We thank the referee for their helpful comments.

This research has made use of the NASA Exoplanet Archive, which is operated by the California Institute of Technology, under contract with the National Aeronautics and Space Administration under the Exoplanet Exploration Program.

\bibliography{references}

\end{document}